# Heat load measurements for the PIP-II pHB650 cryomodule

D. Porwisiak[1,2*], M.J. White[1], V. Roger[1], J. Bernardini[1], B.J. Hansen[1], J.P. Holzbauer[1], J. Makara[1], J. Ozelis[1], D. Passarelli[1], S. Yoon[1], J. Subedi[1], S. Ranpariya[1], V. Patel[1] and J. Dong[1]

[1] Applied Physics and Superconducting Technology Directorate, Fermi National Accelerator Laboratory, Batavia Il, USA
[2] Faculty of Mechanical and Power Engineering, Wroclaw University of Science and Technology, Wroclaw, Poland

*E-mail: dporwisi@fnal.gov

**Abstract.** Phase-3 testing of the pHB650 cryomodule at the PIP-II Injector Test Facility was conducted to evaluate the effectiveness of heat load mitigations performed after earlier phases of testing and to continue pinpointing any sources of unexpectedly high heat loads.. The programme measured HTTS, LTTS, and 2 K isothermal/non-isothermal loads under "standard", "linac", and "simulated dynamic" operating modes, recording data both inside the cryomodule and across the bayonet can circuits. Thermal-acoustic oscillations were eliminated by replacing the original G10 cooldown-valve stem with a stainless-steel stem fitted with wipers. A newly developed Python script automated acquisition of ACNET data, performed real-time heat-load calculations, and generated plots and tables that were posted to the electronic logbook within minutes, vastly reducing manual effort and accelerating feedback between SRF and cryogenics teams. Analysis showed that JT heat-exchanger effectiveness and temperature stratification in the two-phase and relief piping strongly influence the observed loads and helped isolate sources of excess heat. The campaign demonstrates that rigorous pre-test planning, real-time diagnostics, and automated reporting can improve both accuracy and efficiency, providing a template for future PIP-II cryomodule tests and for implementing targeted heat-load mitigations.

## 1. Introduction

The Proton Improvement Plan II (PIP-II) cryogenic plant currently being installed will provide the refrigeration to support superfluid helium baths at 2 K along a string of superconducting radio-frequency (SRF) cryomodules. Design parameters of PIP-II cryoplant are summarised in Table 1.

The PIP2IT (PIP-II Injection Test) cryogenic distribution system was originally designed and commissioned to cool and supply superfluid helium to the half-wave resonator (HWR) and pSSR1 cryomodules during beam-on tests of the PIP-II injection linac. Upon successful demonstration of accelerator operation, PIP2IT was reconfigured as a flexible cryomodule test stand for PIP-II cryomodules. The half-wave resonator (HWR) cryomodule featured a unique bayonet arrangement, so an adapter transfer line was fabricated to convert that bayonet pattern into the standard interface used by HB650 (High Beta 650 MHz) and LB650 (Low Beta 650 MHz) cryomodules. All subsequent SSR1 (Single Spoke Resonator 1) and SSR2 (Single Spoke Resonator 2) modules share the common PIP-II bayonet geometry and require no such adapter transfer line. To enable accurate measurement of static and dynamic High Temperature Thermal Shield (HTTS)

**Table 1.** Design parameters of PIP-II cryoplant [1].

| Circuit | Supply Temp, K | Supply Pressure, bar a | Max Return Temp, K | Min Return Pressure, bar a | Acceptance Test Mass Flow Rate, g/s | Acceptance Test Capacity, W |
|---|---|---|---|---|---|---|
| HTTS | 40 | 18 | 80 | 17.7 | 53 | 10,680 |
| LTTS | 4.5 | 2.7 | 9 | 2.4 | 35 | 1,538 |
| 2K | 4.5 | 2.7 | 4.1 | 0.027 | 132 | 3,104 |

and Low Temperature Thermal Shield (LTTS and 4.5 K cavity) heat loads—which had previously been unattainable due to lack of precise flow data—a flowmeter was installed on the cooldown line to compressor suction. The HB650/LB650 test slot employs 4.5 K supply and 2 K return U-tubes similar in size and design to those in the PIP-II cryogenic distribution system, as does the SSR1/SSR2 slot. Historically, HTTS supply temperatures at PIP2IT were allowed to float below 40 K, producing slow oscillations; a closed-loop heater control on the HTTS supply has now been implemented so that "linac-condition" tests use a stable supply just above 40 K, in contrast to earlier "standard-condition" results (Table 2.). Although the concentric transfer line between the 4.5 K feed and 2 K return should theoretically precool the incoming fluid, the high non-isothermal load on the 2 K circuit typically reheats the 4.5 K supply—except under high dynamic power tests, when increased mass flow lowers the VLP return temperature. The 5K transfer line turnaround valve bullet has now been replaced with a Cv value three times greater, which should increase the

**Table 2.** Operating conditions: "standard" and "linac". Bath liquid level equal to 87% at 23 torr.

| Parameter | Standard conditions, W | Linac conditions, W |
|---|---|---|
| HTTS CM Inlet Temperature, K | 33.9 | 46.5 |
| HTTS CM Outlet Temperature, K | 44.2 | 68.4 |
| LTTS CM Inlet Temperature, K | 5.55 | 5.15 |
| LTTS CM Outlet Temperature, K | 6.47 | 5.90 |

5 K turnaround mass flow rate and reduce the effects of heat transfer across the concentric section of the transfer line. PIP2IT pHB650 configuration was described in [2].

*1.1 Importance of heat loads testing in cryomodules*

The static 2K circuit heat loads measured at PIP2IT thus far have come in higher than expected, which is of concern due to the fixed PIP-II cryoplant capacity. However, thus far only the HWR, pSSR1, and pHB650 cryomodules have been tested at PIP2IT. If the causes of the higher than expected can be understood there remains some time to put heat load mitigations in place and prevent future cryomodule static 2K heat loads from continuing to reduce the available PIP-II cryoplant capacity margin.

## 2. Overview of previous testing phases

*2.1 Summary of Phase 1 and Phase 2 testing*

Phase 1 testing was conducted with the cryomodule in its baseline configuration: the cooldown and Joule-Thomson valves used stainless-steel stems without wipers, and the HTTS and LTTS supply temperatures were allowed to float with the load on the Superfluid Cryogenic Plant at CMTF (Cryomodule Test Facility). Because the HTTS supply was floating below 40K, the resulting temperature profiles diverged from those expected in PIP-II linac service; nevertheless, this "standard-conditions" measurement sequence has been preserved in testing for direct comparison in later campaigns. In Phase 2, several refinements were introduced. Improved control on both HTTS and LTTS headers stabilized their temperatures at set-points derived from the PIP-II cryoplant design, yielding a closer approximation to linac conditions that has since been adopted for all "linac-conditions" measurements. Hardware upgrades complemented these operational changes: additional radiation shielding was installed around the power couplers, and the cooldown and Joule-Thomson valves were rebuilt with lower thermal conductivity G10 stems—still without wipers—to mitigate thermal acoustic oscillations. Although HTTS and LTTS heat loads were not recorded for the HWR module and further tests on SSR2, and LB650 are pending, extrapolated cryomodule data permits estimates of the target linac conditions temperature distribution. The comparative results from Phases 1 and 2 are summarized in Table 3.

**Table 3.** Phase 1 and 2 testing results at Standard and Linac conditions

|  | LTTS | HTTS | 2K isothermal | 2K total |
|---|---|---|---|---|
| Design value, W | 26.3 | 144 | 9.6 | 14.2 |
| Phase 1 measured values, W | 29.7 ($T_{in}$ = 5.51 K $T_{out}$ = 6.90 K) | 260 ($T_{in}$ = 34.3 K $T_{out}$ = 45.4 K) | 50.8 | 79.8 |
| Phase 2 measured values at Standard conditions, W | 23.3 ($T_{in}$ = 5.55 K; $T_{out}$ = 6.47 K) | 240 ($T_{in}$ = 33.9 K; $T_{out}$ = 44.2 K) | 38.2 | 57.3 |
| Phase 2 measured values at Linac conditions, W | 25.9 ($T_{in}$ = 5.15 K; $T_{out}$ = 5.90 K) | 203 ($T_{in}$ = 46.5 K; $T_{out}$ = 68.4 K) | 38.2 | 61.5 |

## 2.2 pSSR1 Heat loads testing

The pSSR1 cryomodule was retested to take advantage of flowmeters installed for measuring pHB650 HTTS and LTTS flow rates. LTTS heat load measurements exceeded design values more than 3 times. On the other hand, HTTS measured heat loads were slightly lower than expected. Cryomodule 2 K static heat loads obtained during this testing were over 3 times higher than design values. Flow diagram of pSSR1 cryomodule is presented on Figure 1. Results of heat load measurements are presented in Table 3.

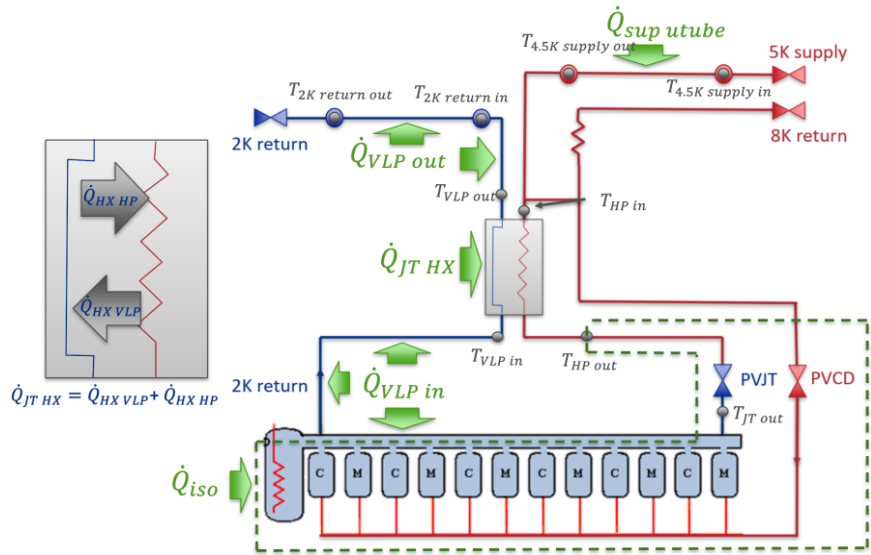

**Figure 1.** Flow diagram of pSSR1 cryomodules with marked measured heat loads.

**Table 3.** pSSR1 static heat loads results

|  | LTTS | HTTS | 2-Phase Pipe & JT HX VLP Inlet | JT HX | JT HX VLP Outlet & U-tube | 2K isotherm | 2K total sum | 2K total energy balance |
|---|---|---|---|---|---|---|---|---|
| Design value, W | 21.9 | 274 | 0.3 (CM) + 3.9 (U-tube)[a] | | | 13 | 17.2 | 18.5 |
| Standard condition, W | 71 | 263 | 8.2 ($T_{VLP,in}$ = 2.80K) | -9.8 ($T_{VLP,out}$ = 3.01K & 3.34K) | 21.4 ($T_{2K\_ret\_out}$ = 5.43K) | 32.3 | 52.1 | 51.4 |
| Linac condition, W | 94 | 263 | 9.7 ($T_{VLP,in}$ = 2.86K) | -1.9 ($T_{VLP,out}$ = 4.11K & 4.71K) | 13.4 ($T_{2K\_ret\_out}$ = 5.69K) | 35 | 56.2 | 56.8 |

[a] This sum applies to total 2K non-isothermal heat load

*2.3 Implications for PIP-II cryoplant capacity*

The fixed design of the Joule–Thomson heat exchangers (JT HXs)—with remaining units already on order—precludes retroactive modification of their nominal effectiveness ($\varepsilon$) without compelling performance data. At present, neither the JT HX performance characterization nor the aggregate cryomodule heat-load measurements are of sufficient precision to justify JT HX redesign. Consequently, the operational envelope of the PIP-II cryogenic plant must be managed through system-level adjustments rather than component changes.

One such adjustment is the deliberate boiling of 2 K liquid inventory within the cryoplant, cryomodules or the tunnel transfer line "turnaround" can. By vaporizing a controlled fraction of the 2 K helium bath, the cold-compressor inlet temperature can be reduced toward the 4.1 K target via mixing. However, this strategy carries a direct penalty: each gram per second of deliberate boil-off constitutes unavailable refrigeration capacity for isothermal (static + dynamic) loads on the SRF cavities.

The PIP-II cryoplant is designed for 132 g/s mass flow (see Table 1.) when the cold-compressor inlet is held at 4.1 K; any rise above this cold compressor inlet temperature further constrains allowable mass flow rate. Consequently, elevated non-isothermal 2 K heat loads that result in return temperatures greater than 4.1 K reduce the cryoplant's isothermal capacity.

## 3. Methodology of Phase 3 Heat Load Testing

*3.1 Description of test setup, procedures and improvements for Phase 3 testing*

Phase 3 heat-load measurements at PIP2IT showed progress in addressing the unexpectedly high loads observed during earlier campaigns—an issue of critical importance given the fixed capacity of the PIP-II cryoplant. Only three cryomodule types (HWR, pSSR1 and pHB650) had been tested to date; by isolating the root causes of the excess heat loads, design or procedural changes can still be implemented for future cryomodules before they enter the test stand to mitigate further erosion of cryoplant capacity margin.

In Phase 3 the test configuration was further refined to improve both reliability and diagnostic precision. A comprehensive inspection confirmed that the cryomodule sustained no mechanical or thermal damage during round-trip shipment to UKRI. A check valve was installed on the relief line, replacing previously installed relief valve. Diagnostic coverage was expanded by adding several high-accuracy temperature sensors together with a dedicated 2 K helium flow-meter, enabling finer resolution of heat-load balance and stability analyses. Finally, the cooldown valve stem was cold-swapped: the G10 stem was replaced with a stainless-steel stem with wipers, a measure proven to suppress thermal-acoustic oscillations and minimise parasitic heat input.

A major advance for Phase 3 was the development of an automated Python-based workflow that retrieves raw sensor data from ACNET (Accelerator Control Network), computes static and dynamic heat loads, and generates comprehensive PDF reports—complete with plots and summary tables—immediately upon test completion. This streamlined approach, replaced the labor-intensive manual extraction and spreadsheet entry of prior campaigns, enabling far more rapid and thorough analysis. All Phase 3 data and reports were promptly archived in the electronic logbook, ensuring full traceability.

In parallel, a real-time heat-load calculation capability was implemented by invoking the HEPAK thermophysical library via EPICS. Although Phase 3 initially incorporated only basic HTTS and LTTS heat-load channels, the EPICS-computed values agreed closely with the Python post-processing results. This real-time feedback allows the cryogenic operator to monitor measurement stability continuously—superseding the former practice of arbitrarily long "steady-state" intervals—and to confirm, directly on an HMI display, when temperature, pressure and flow signals have converged to a reliable heat-load value [3].

Another procedural enhancement was the pre-definition of a detailed test matrix prior to cooldown: for each measurement point, target temperatures, stability criteria and minimum hold times were established in advance. While operational interruptions—such as cryoplant upsets, parallel cryomodules or RF testing, and scheduling constraints—occasionally prevented attainment of every aspirational target, the Python-generated reports enabled rapid assessment of data quality.

Finally, the Phase 3 campaign adopted a comprehensive reporting format inspired by best practices at ESS. Rather than simple spreadsheets, this document furnishes a full narrative of test objectives, methodologies, data analysis and results, supplemented by lessons learned and recommended best practices. This structured framework will serve as the template for all future PIP2IT cryomodule testing campaigns, ensuring clarity, reproducibility and continuous improvement.

Following tests we performed were:

a. Standard conditions test – during these tests, particular attention was given to ensuring that the parameters conformed to those specified in the Table 4 'Standard conditions'. In addition to cavity circuit, also HTTS and LTTS heat loads were measured. Parameters were kept for much longer periods of time to make sure system is stable and collect most reliable data (overnight or over the weekend).

b. Linac conditions test – during these tests, particular attention was given to ensuring that the parameters conformed to those specified in the Table 4 'Linac conditions'. In addition to cavity circuit, also HTTS and LTTS heat loads were measured. Parameters were kept for much longer periods of time to make sure system is stable and collect most reliable data (overnight or over the weekend).

c. dP/dT heater calibration test – measurements performed for 0 W, 5 W, 10 W, 20 W and 30 W. For each heater power set-point, the system pressure $P(t)$ is recorded over time and each trace is fitted with a straight line. The slope of that fit, $dP/dt$, gives the pressure rise rate per watt of heater input. Plotting these $dP/dt$ values versus heater power yields a calibration curve which, when inverted and extrapolated to zero power, provides the static heat load 37.19 W for pHB650 cryomodule.

d. 4K with different mass flow rates on LTTS line – data was collected for 2 different mass flow rates flowing through LTTS line to check the influence of conduction heat load to the cavity circuit. Difference in static heat loads during this test was ~0.5 W which is less than uncertainty margin.

e. Fake unit test – Data collected for different cavity powers up to 137 W to check heat load under operating accelerator conditions. After reaching 137 W power was gradually decreased to check hysteresis. The results in figure 2 show that the estimated static isothermal heat loads are up to 5 W higher while reducing heater power compared to when increasing the heater power. Other phenomenon observed during this test was change of temperature profile in two-phase pipe further described in 4.3.

*3.2 Measurements techniques and instrumentation*

Accurate determination of static and dynamic heat loads in the PIP2IT cryogenic distribution system relies on redundant, high-precision instrumentation and automated data acquisition. Temperatures along the 4.5 K supply, 2 K return, HTTS and LTTS headers are monitored by calibrated Cernox™ and silicon diode sensors, with typical uncertainties below ±10 mK. Helium mass-flow is measured on the return side using a thermal-mass flowmeter with uncertainty ± 0.5% full scale (compressor suction line).

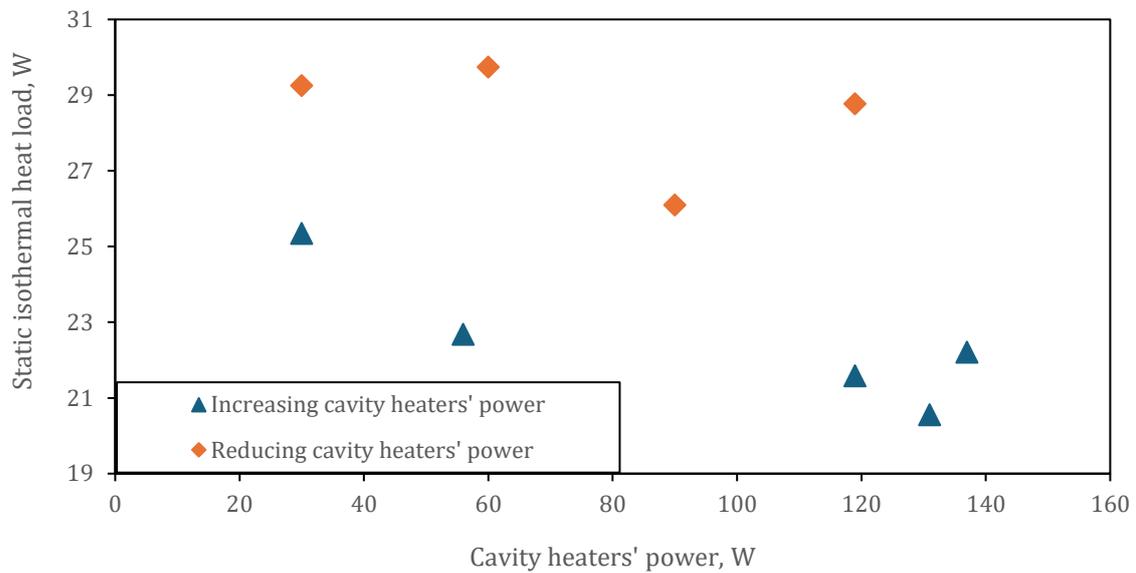

**Figure 2.** Static isothermal heat load during fake unit test.

## 4. Key observations and results

### 4.1 Measured heat loads

**Table 5.** Phase 3 static heat loads results

|  | LTTS | HTTS | 2K isothermal | 2K total[a] |
|---|---|---|---|---|
| Design value, W | 26.3 | 144 | 9.6 | 14.2 |
| Standard condition, CD valve with G10 stem | 19.5 ($T_{in}$ =5.4 K $T_{out}$ =6.0 K) | 239 ($T_{in}$ =34.4 K $T_{out}$ =44.8 K) | 31.7 | 48.9 |
| Linac condition, CD valve with G10 stem | 24.2 ($T_{in}$ =5.3 K $T_{out}$ = 5.8 K) | 207 ($T_{in}$ = 49.1 K $T_{out}$ = 68.6 K) | 30.9 | 48.5 |
| Linac condition, CD valve with SS stem w/ wipers | 26.5 ($T_{in}$ = 5.4K $T_{out}$ = 6.1K) | NA | 29.2 | 47.2 |

[a] Including u-tubes heat loads.

Figure 3 illustrates the flow diagram of pHB650 heat loads measurements. Diagram does not include HTTS circuit. Heat loads measurements of pHB650 is summarised in Tables 5 and 6

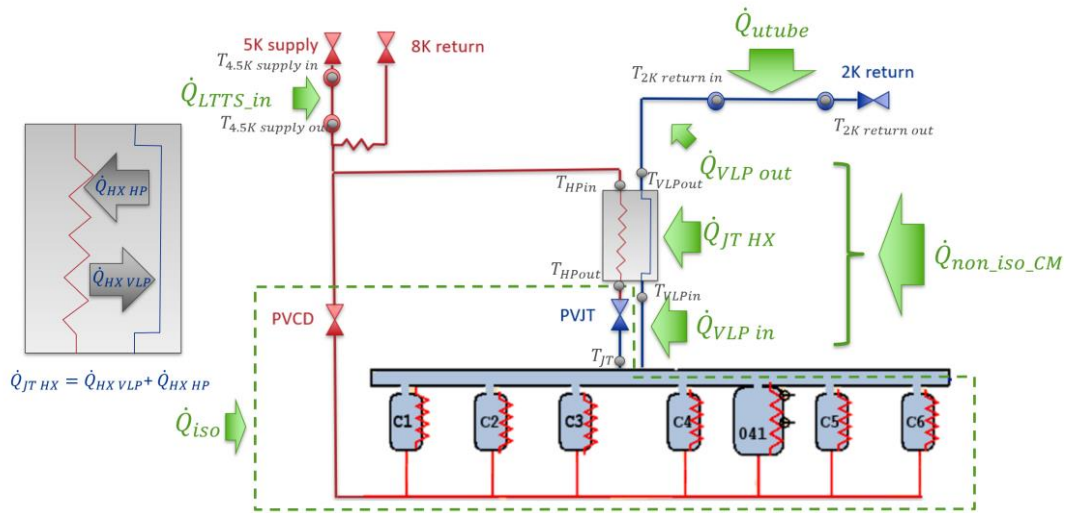

**Figure 3.** Flow diagram of pHB650 cryomodules with marked measured heat loads.

**Table 6.** Phase 3 static non-isothermal heat loads of 2K circuit

|  | JT HX VLP Inlet | JT HX | VLP Outlet & 2K Return U-tube | 4.5 K supply U-tube |
|---|---|---|---|---|
| Design value, W | 0.3[a] |  | 3.9 | 1.3 |
| Standard condition, CD valve with G10 stem, W | 3.2 | 3.0 | 8.7 | 1.5 |
| Linac condition, CD valve with G10 stem, W | 3.4 | 3.2 | 9.2 | 0.7 |
| Linac condition, CD valve with SS stem w/ wipers, W | 3.2 | 3.3 | 9.3 | 1.4 |

[a] This value is the sum of JT HX VLP Inlet + JT HX

*4.2 JT heat exchanger performance*

The significant number of tests allows for the analysis of the heat exchanger and the comparison of experimental results with those simulated by the manufacturer. To compare the performance of the heat exchanger during different tests, the results were presented in the form of a NTU – efficiency characteristic (Figure 4.). On the graph all 2K tests were included. For low NTU values efficiency of the heat exchanger is lower than predicted by manufacturer, while for high NTU effectiveness is higher than expected. However the results are too scattered to determine if the heat exchanger meets required performance criteria. Data scatter is believed to be primarily caused by the placement of surface-mounted temperature sensors on the VLP piping. The VLP sensor locations were updated for future PIP-II cryomodules.

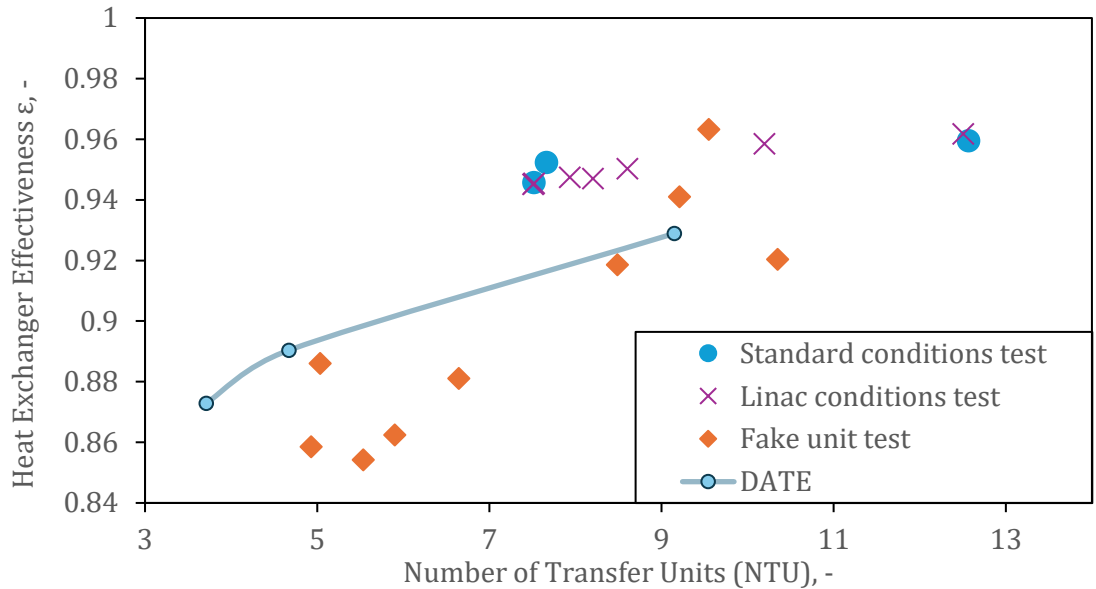

**Figure 4.** JT Heat Exchanger Performance for the PIP-II pHB650 Cryomodule Phase 3 Testing at PIP2IT. DATE stands for calculated data supplied by manufacturer.

*4.3 Temperature distribution in two phase pipe*
Two phase line temperature distribution depends on whether the cannister heater or cavity heaters were used. JT HX VLP inlet temperature was 0.1 K lower using cannister heater than when applying the same heater power using cavity heaters. Additionally different temperature profiles were observed in two phase pipe during fake unit test. For highest values of cavity heaters power top mounted sensors showed lower temperatures than sensors on the side of the pipe (Figure 5).

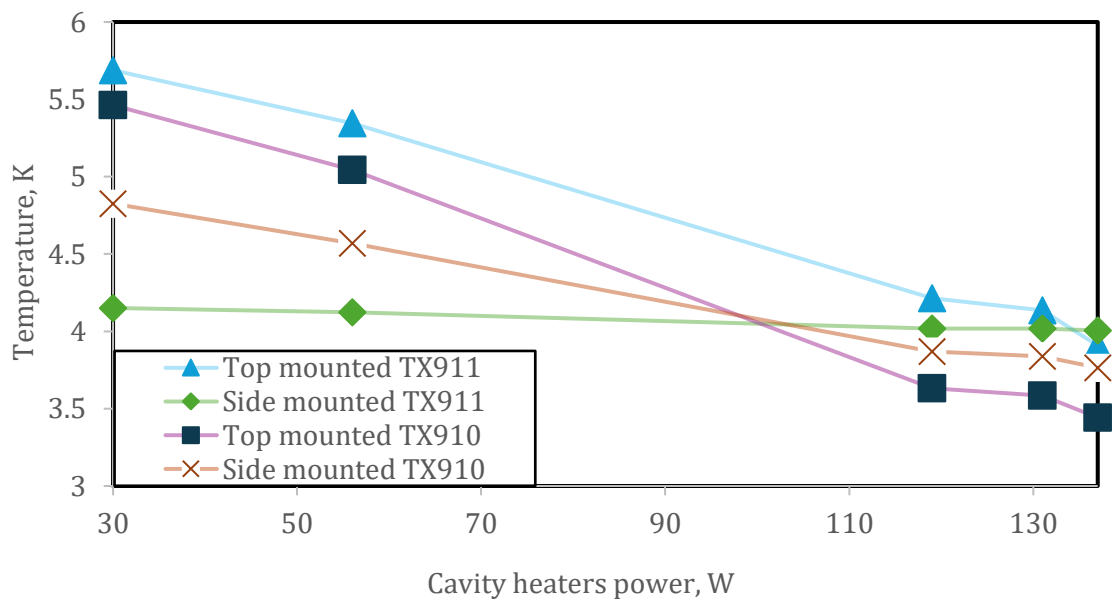

**Figure 5.** Temperature curves for 4 of the sensors mounted on the two phase pipe. TX911 are located between heat exchanger and relief stack and TX910 are located on another side of relief stack.

## 5. Recommendations for future cryomodule tests

Two design issues were identified and subsequently addressed for the production cryomodules [4]. The high heat loads observed on the HTTS were determined to be solely attributable to insufficient MLI surrounding the G11 tube of the support posts. Calculations revealed direct radiation from the vacuum vessel to the internal components of the cold mass. This phenomenon accounts for the high LTTS heat loads and contributes to the elevated 2K heat loads. To mitigate this, the thermal shield and MLI have been redesigned for future cryomodules. Moreover, additional modifications and lessons learned have been incorporated to further reduce the 2K heat loads. However, a residual 10W on the 2K heat loads remained unexplained by predictive calculations

## 6. Summary

The HTTS, LTTS, and 2K circuit static heat loads were comprehensively measured for the pSSR1 and pHB650 cryomodules. Although the heat loads were higher than expected progress was made in reducing the pHB650 heat loads during each of the 3 phases of testing. Based on the measured performance additional design and fabrication changes are being made to future PIP-II cryomodules to reduce static heat loads further and ensure that the PIP-II cryoplant capacity is not exceeded.

## Acknowledgments
This work was produced by Fermi Forward Discovery Group, LLC under Contract No. 89243024CSC000002 with the U.S. Department of Energy, Office of Science, Office of High Energy Physics.